\documentclass[11pt, reqno]{amsart}
\usepackage{graphicx,amsmath,amssymb,epsfig}
\usepackage{amsfonts}

\theoremstyle{plain}

\numberwithin{equation}{section}

\begin{document}
\title[VaR at risk]{The VaR at risk}
\author{Alfred Galichon$^\dag$}
\date{First version is June 3, 2008. The present version is of January 14, 2010.\\
$^\dag$Correspondence address: D\'{e}partement d'Economie, Ecole polytechnique,
91128 Palaiseau, France. The author thanks Jean-Charles Rochet and Ren\'{e} A%
\"{\i}d for encouragements and stimulating discussions. Support from Chaire
EDF-Calyon \textquotedblleft Finance and D\'{e}veloppement
Durable,\textquotedblright\ Chaire Soci\'{e}t\'{e} G\'{e}n\'{e}rale
\textquotedblleft Risques Financiers\textquotedblright\ and Chaire Axa
\textquotedblleft Assurance et Risques Majeurs\textquotedblright \thinspace\
and FiME, Laboratoire de Finance des March\'{e}s de l'Energie
(www.fime-lab.org) is gratefully acknowledged.}

\begin{abstract}
I show that the structure of the firm is not neutral with respect to
regulatory capital budgeted under rules which are based on the
Value-at-Risk. Indeed, when a holding company has the liberty to divide its
risk into as many subsidiaries as needed, and when the subsidiaries are
subject to capital requirements according to the Value-at-Risk budgeting
rule, then there is an optimal way to divide risk which is such that the
total amount of capital to be budgeted by the shareholder is zero. This
result may lead to regulatory arbitrage by some firms.
\end{abstract}

\maketitle

\noindent


{\footnotesize \textbf{MSC 2000 subject classification}: 91B06, 91B30, 90C08 %
\vskip50pt }



\section{Introduction} The single most used measure of financial risk is
undoubtedly the Value-at-Risk (VaR). The VaR at level 95\% is defined as the
minimal amount of capital which is required to cover the losses in 95\% of
cases. In statistical terms, the value-at-risk is the quantile of level $%
\alpha $ of the losses, namely%
\begin{equation*}
VaR_{\alpha }\left( X\right) =\inf \left\{ x:\Pr \left( X\leq x\right)
>\alpha \right\} .
\end{equation*}%
(note that unlike the most widely adopted convention in the literature, we
chose to count positively an effective loss).

The widespread popularity of the Value-at-Risk is due to its adoption as a
the ``1st pillar'' in the Basle II agreements. Despite its widespread use
and simplicity, the Value-at-Risk is highly criticized among academics. The
literature on risk measures classically defines a set of axioms which
satisfactory risk measures should satisfy (see \cite{ADEH:99}). Among these, 
\textit{subadditivity}: a risk measure $\rho $ should satisfy $\rho \left(
X+Y\right) \leq \rho \left( X\right) +\rho \left( Y\right) $. While this
axiom is widely accepted in the academic community, it is perhaps ironical
that VaR fails to satisfy this subadditivity axiom as it has been widely
documented (we come back to that below).

The subadditivity axiom is generally motivated by a loose invocation of risk
aversion, or preference for diversification. In this note I propose a quite
different argument to motivate the importance of the subadditivity axiom.
Assuming the value-at-risk is used to budget regulatory capital
requirements, and assuming that the managers have an incentive to minimize
these capital requirements, I show that the lack of the subadditivity
property induces the possibility for the management to optimally divide
their risk in order to minimize their budgeted capital: the structure of the
firm is not neutral to the aggregated capital requirement. More precisely, I
show that for any level of the value-at-risk there is a division of the risk
which sets the aggregated capital requirement to zero.

\section{The problem} Consider a trading floor which is organized into $N$
various desks. For each trading desk $i=1,...,N$, call $X_{i}$ the random
variable of the contingent loss of trading floor $i$. The total random loss
of the trading floor is $X=\sum_{i=1}^{N}X_{i}$, we suppose that this amount
is bounded: $X\in \left[ 0,M\right] $ almost surely.

We suppose that each desk budgets a regulatory capital equal to its VaR at
level $\alpha \in \left( 0,1\right) $, $VaR_{\alpha }\left( X_{i}\right) $.
Consequently the total amount of regulatory capital which the management
needs to budget is $\sum_{i=1}^{N}VaR_{\alpha }\left( X_{i}\right) $. It is
therefore easy to formulate the manager's problem:%
\begin{equation}
\inf \sum_{i=1}^{N}VaR_{\alpha }\left( X_{i}\right) \text{s.t. }%
\sum_{i=1}^{N}X_{i}=X\text{ a.s., and }VaR_{\alpha }\left( X_{i}\right) \geq
0.  \label{mp}
\end{equation}

As the management has full control over the structure of the firm, it
results that it has the choice over $N$ and over the random variables $%
\left( X_{1},...,X_{N}\right) $ which satisfies the constraints.

Note that the economical risk which the trading floor bears is $%
X=\sum_{i=1}^{N}X_{i}$, and thus the regulatory capital to be budgeted
should be $VaR_{\alpha }\left( X\right) >0$. However, I shall show that,
under mild assumptions, there is a structure of the firm such that the
capital budgeted under the rule above is zero. The assumption needed is the
following:

\textbf{Assumption.} \textit{The distribution of }$X$\textit{\ is absolutely
continuous with respect to the Lebesgue measure.}

As we shall see in Point 6. in the discussion below this assumption can in
fact be removed.

\textbf{The optimal structure.} In that case, there exists a sequence of
real numbers $x_{0}=0<x_{1}<...<x_{N}=M$ such that $\Pr \left( X\in
(x_{i},x_{i+1})\right) <1-\alpha $ for all $i=1,...,N-1$.

We are therefore going to consider the \textit{digital options}\footnote{%
this is a slight abuse of terminology as a digital option would more
correctly characterize $x_i1\left\{ X\in (x_{i},x_{i+1})\right\}$, but both
expressions are close when $x_i$ and $x_{i+1}$ are close.}%
\begin{equation*}
X_{i}=X1\left\{ X\in (x_{i},x_{i+1})\right\}
\end{equation*}%
(note that these options can be approximated by linear combinations of
standard calls and puts, or \textit{butterfly options}). We have%
\begin{equation*}
\sum_{i=1}^{N}X_{i}=X.
\end{equation*}%
As we have $\Pr \left( X_{i}>0\right) <1-\alpha $, it follows that 
\begin{equation*}
VaR_{\alpha }\left( X_{i}\right) =0.
\end{equation*}

Therefore the capital to be budgeted under this structure of the firm is the
sum of the capital to be budgeted for the different trading desks, which is
zero.

\section{Discussion}

1. This example shows that it is possible to tear down the risk into small
pieces which are indetectable by the Value-at-Risk. Quite obviously, it is
far too stylized to aim at modelling any of the real-life events that took
place on the credit market in 2007-2008. Instead its ambition is to
providing an alternative illustration of a major flaw in the Value-at-Risk
-- namely the fact that it is insensitive to the seriousness of the events
in the tail. It takes into account the probability that the threshold point $%
VaR_{\alpha }$ be reached, not how serious the losses are beyond that point.

2. Since the seminar work of Modigliani-Miller, a vast literature in
corporate finance has discussed whether the structure of the capital of the
firm does impact or not the value of the firm for the stakeholders. In
Modigliani-Miller's result the value of the firm depends linearly on the
risky capital claims, thus yielding to the celebrated structural neutrality
result. When one introduces nonlinear tax on risky capital (such as the
Value-at-Risk in the present setting), the result is reversed, and one has
to conclude to the non-neutrality of the capital structure.

3. An obvious consequence of this paper's point is that the VaR is not
additive. Note that there are cases where VaR is additive, notably when all
the $X_{i}$'s are \textit{comonotonic}. Comonotonicity indeeds turns out to
be the regulator's best case which justifies the \emph{comonotonic additivity%
} axiom put forward a large literature: see \cite{Kusuoka:2001}.

4. The requirement that a measure of risk should be immune to regulatory
arbitrage has led \cite{GH08} to define the axiom of \textit{strong coherence%
}, which is a natural requirement so that the structure of the firm be
neutral to risk measurement. In that paper, strong coherence is shown to be
equivalent to the classical risk measures axioms in \cite{ADEH:99}.

5. The assumption that the distribution of the risk $X$ be absolutely
continuous was necessary for the argument\ presented above to work, and
there are connections to be explored with the theory of portfolio
diversification under thick-tailedness, see \cite{Ibragimov:2004}.

6. However when the distribution of the risk has atoms of mass greater than $%
\alpha $, it is still possible decrease the regulatory capital to zero by 
\emph{randomizing}, i.e. randomly assigning the risk to one out of several
subsidiaries which shall then take on the total loss. The idea is the
following: consider $N\geq 1/\alpha $ and incorporate a number $N$ of
subsidiaries indexed by $i\in \left\{ 1,...,N\right\} $. When the parent
company faces loss $X$, an index $i$ is drawn randomly, uniformly in $%
\left\{ 1,...,N\right\} $, and the parent company turns to subsidiary $i$ to
cover all off the company's loss $X$. The parent company has no residual
risk: in it in any case insured by one of its subsidiaries. All the
subsidiaries have a probability $1/N$ which is less than $\alpha $ of being
called, so each of their $VaR_{\alpha }$ is zero.

7. When there is an upper bound to the number $N$ of subsidiaries that can
be created, the problem of the minimal capital cost of risk $X$ given by (%
\ref{mp}) is still open.

\section{Possible extension} It would be interesting to consider the above
problem with overhead costs, namely when dividing the firm into several
desks is costly. The problem would then become%
\begin{equation*}
\inf \sum_{i=1}^{N}VaR_{\alpha }\left( X_{i}\right) +\alpha \left( N\right) 
\text{ s.t. }\sum_{i=1}^{N}X_{i}=X\text{ a.s., and }VaR_{\alpha }\left(
X_{i}\right) \geq 0.
\end{equation*}%
where $\alpha \left( N\right) $ is increasing with $N$ and can be
interepreted as an \emph{a priori penalization of the firm's complexity} by
the regulator. The problem of determining an optimal $\alpha \left( N\right) 
$ from the point of view of the regulator in a properly defined setting is
of interest.

\textit{$^\dag$ \'Ecole polytechnique, Department of Economics. E-mail:
alfred.galichon@polytechnique.edu} 

\end{document}